\documentclass[journal=jacsat,manuscript=article]{achemso}

\usepackage{appendix}
\usepackage{color}
\usepackage[version=3]{mhchem} 

\author{Mahdi Safari}
\affiliation[University of Toronto]
{Department of Electrical Engineering, University of Toronto, Toronto, Canada}
\author{Nazir P. Kherani}
\affiliation[University of Toronto]
{Department of Electrical Engineering, University of Toronto, Toronto, Canada}
\altaffiliation{Department of Material Science and Engineering, University of Toronto, Toronto, Canada}
\author{Geroge V. Eleftheriades}
\email{gelefth@ece.utoronto.ca}
\phone{}
\fax{}
\affiliation[University of Toronto]
{Department of Electrical Engineering, University of Toronto, Toronto, Canada}

\date{\today}

\title[An \textsf{achemso} demo]
  {Multi-functional Metasurface:
Visibly and RF Transparent, NIR Control and Low Thermal Emissivity}

\abbreviations{IR,NMR,UV}

\begin{document}


\begin{abstract}
Rapid advances in metamaterial technology are enabling the engineering of wave-matter interactions heretofore not realized and functionalities with potentially far-reaching implications for major challenges in the fields of energy conservation and radio frequency (RF) communication. We propose a visibly and RF transparent composite metasurface utilizing dielectric-metal spectrally selective coatings with high NIR control and low thermal emissivity, thus achieving a multi-functional metasurface capable of enhancing 5G communication efficiency and exhibiting energy conservation features. The proposed meta-glass yields 92\% peak RF transmission at 30 GHz which corresponds to 20\% and 90\% enhancement when compared to plain glass and low-emissive glass substrates. This meta-glass possesses 86\% peak optical transparency at $\lambda=550 nm$, $>$60\% near-IR reflection, and $>$ 80\% mid-IR reflection which corresponds to $\approx$ 0.2 thermal emissivity. The proposed metasurface design is highly flexible and can be tuned to operate over different frequency ranges owing to its frequency scalability. This study provides a better alternative using earth-abundant materials compared to low-emissive glass based on indium tin oxide (ITO) while boosting the efficiency of 5G communication amenable to window systems demanding simultaneous functionalities for emergent smart/energy-efficient buildings/cities and autonomous transportation applications.
\end{abstract}


\maketitle


\section{\label{sec:level1}Introduction}
The growing recognition of the need for sustainable energy practices over the last several decades has drawn research attention to developing window technologies that provide desired functionalities and yet promote energy conservation. The extensive use of glass windows in modern architecture and smart city environments underscores the importance of transparent low thermal emissive coatings 
vis-à-vis energy consumption. In this regard, low emissive coatings have been extensively studied and include conductive oxides such as indium thin oxide (ITO), aluminum zinc oxide (AZO), vanadium oxide (VO$_2$), and dielectric-metal-dielectric (DMD) layer stacks\cite {ding2017silver,ando2001moisture,yuan2013influence,liu2017achieving,alonso2017ito,jelle2015low,xu2019optical,ko2018ultrasmooth}. 
While considerable development of various low-emissive coatings has been realized, the use of highly conductive materials for near-infrared (NIR) control results in the blockage of RF signals which makes the implementation of these devices in the smart city environment dubious\cite{green2019optically}. To mitigate this drawback, recent studies have suggested sub-wavelength patterning of low-emissive films to preclude barrier properties of uniform silver films in the low RF frequency range by virtue of disconnected repeating pattern of Ag islands \cite{liu2017achieving}. However, with the advances in 5G telecommunication there is a pressing need to develop versatile and flexible designs capable of operating in the higher frequency range extending to 100 GHz \cite{Safari2020optically,green2019optically}.

With the introduction of metasurfaces, 2D periodic structures comprising sub-wavelength inclusions capable of engineering unconventional optical and radio frequency (RF) properties, the possibility of designing novel materials such as invisibility cloaks has been extensively studied \cite{ni2015ultrathin,zhang2008plasmon,safari2017cylindrical,safari2019illusion}. Other reports detail parity-time symmetry structures with balanced loss and gain resulting in exotic EM properties \cite{safari2018shadow,miri2019exceptional}. Yet other investigations report on plasmonic \cite{zhang2008plasmon} and magnetic \cite{shulga2018magnetically} induced transparency. In a recent study, a thick layer of densely packed silver nanoparticles was proposed to achieve wideband visible transparency \cite{palmer2019extraordinarily}. These proposed concepts while presenting various advantages in the visible, they are largely theoretical and lack experimental demonstration. Further, in a number of these designs the presence of active elements or the complexity of the designed structures renders their fabrication and their experimental verification impractical considering the required material  properties, if not unfeasible. 

In contrast, over the last few years there has been a focus on the development and fabrication of transparent absorptive and non-reflective metasurfaces based on thin transparent conductive films \cite{zhao2018optically,hu2017optically,chen2017coding,jang2014transparent,jing2018optically,zhang2019flexible,ruan2019optical,xu2019optical,song2018transparent,faniayeu2017highly}. These metasurfaces utilize ITO and DMD coatings which due to their low electrical conductivity are highly advantageous for RF absorbing and stealth applications, however, their higher loss precludes the design of RF transparent devices in the higher frequency ranges. In this regard, we note that a few recent studies on the design of reflect array antennas employ ITO transparent conductive coatings such that losses are minimized through subwavelength design and use of different unit cell geometries \cite{kocia2016design,dreyer2013design}. However, utilization of versatile metallo-dielectric coatings has received little attention. In light of this and the recognition of the need to overcome the constraints noted above,  the objective of the present study is to design a multifunctional meta-glass which is transparent both in the RF and visible regions, provides control over transmission in the NIR and possesses low thermal emissivity, thus making it a perfect fit for the modern city and 5G applications. The contemplated structures offer two principle advantages: increasing the efficiency of 5G communication via windows and ensuring visible, near-infrared and thermal energy modulation. Moreover, employing metamaterial design concepts permit utilization of its frequency scalability and design flexibility in relation to various applications and frequency bands.

Multilayered spectrally-selective dielectric-metal-dielectric (DMD) coatings, consisting of  Ag (M) and AlN (D) layers, have been developed for visible transparency,  NIR control and low thermal emissivity \cite{ko2018ultrasmooth}. In particular, Ko \textit{et al.} have shown that control over the surface morphology helps increase the electrical conductivity of the layer, thereby reducing ohmic lossess, and thus permitting versatility in the  design of DMD coatings. Herein, we employ these versatile DMD coatings to design a composite metasurface, otherwise referred to as a meta-glass structure or simply meta-glass, for RF transparency at a desired frequency using a multifunctional design which also provides for transparency in the visible, NIR control and low thermal emissivity - in short, properties over 4 frequency ranges of the EM spectrum.  The proposed composite metasurface constitutes two identical nature-inspired honeycomb patterned metasurfaces on each side of a glass substrate. The ability to operate over a wide bandwidth and incident-angle range is of interest in 5G communications. The hexagonal/honeycomb patch based design presented here supersedes our earlier design based on hexagonal/honeycomb grid design \cite{Safari2020optically}; in particular this new design  exceeds in performance in both the RF and optical regions. Moreover, the proposed honeycomb patch meta-glass is capable of NIR control and provides low thermal emissivity. A detailed comparison of the two structures is given in the the supporting information.

This paper is organized as follows. First, we discuss the design of the uniform (not patterned) DMD coatings whereby desired visible transparency, NIR reflection and low-emissivity are attained. Next, we design and examine the meta-glass structure (i.e., the composite metasurface) to achieve high RF transparency centered at 30 GHz and extending over a large range of frequencies and incident angles. Subsequently, we explore the adaptability of the proposed design for various substrate (glass) thicknesses and frequency ranges. We then present the fabrication process and experimental results. Finally, we conclude pondering potential applications and the future of multifunctional metasurfaces and transparent metamaterials. 

\section{Method}

The primary objective of this study is to design a composite metasurface wherein the meta-glass provides desired functionalities in four frequency regions: high transparency in the visible, tunable reflection in the near-infrared, low thermal emissivity (in the mid-infrared), and high RF transparency (see Fig. 1). 
The principle idea is to use a metamaterial design approach whereby the EM properties of glass  can be engineered through the use of identical hexagonal metallic inclusions, that is, honeycomb patches, on both sides of a glass substrate. By changing the geometrical parameters of the honeycomb patch structure we are able to manipulate the induced electrical currents on the surface and tune the EM properties of the composite meta-glass (i.e., honeycomb patch meta-glass). The 2-dimensional symmetry of the meta-glass, based on honeycomb-shaped unit-cells, serves to achieve rotational symmetry and  polarization fidelity (see Fig. 1 a and b). Further, we show that each honeycomb patch patterned metasurface on glass provides surface impedance which makes it possible to achieve unity transmission at high RF frequency (30 GHz) considering the double-stub impedance matching technique \cite{pozar2009microwave}. In addition, we show that it is possible to tune the surface impedance of the honeycomb patch metasurfaces and the operating frequency of the composite meta-glass by virtue of varying the geometrical parameters. 

We begin by discussing the design of a uniform (unpatterned) thermally low-emissive, visibly transparent DMD spectrally selective coating, comprising Ag (M) and AlN (D) layers with high electrical conductivity at RF frequencies. Then we utilize the designed uniform DMD coating to fabricate honeycomb patterned metasurfaces on both sides of the glass so as to achieve RF transparency around 30 GHz - analogous to the transmission line double-stub impedance matching for free-space applications \cite{pozar2009microwave}. 
We also discuss the physics of RF transparency using transmission-line method and impedance matching technique. 
\begin{figure}
    \centering
    \includegraphics[width=15cm]{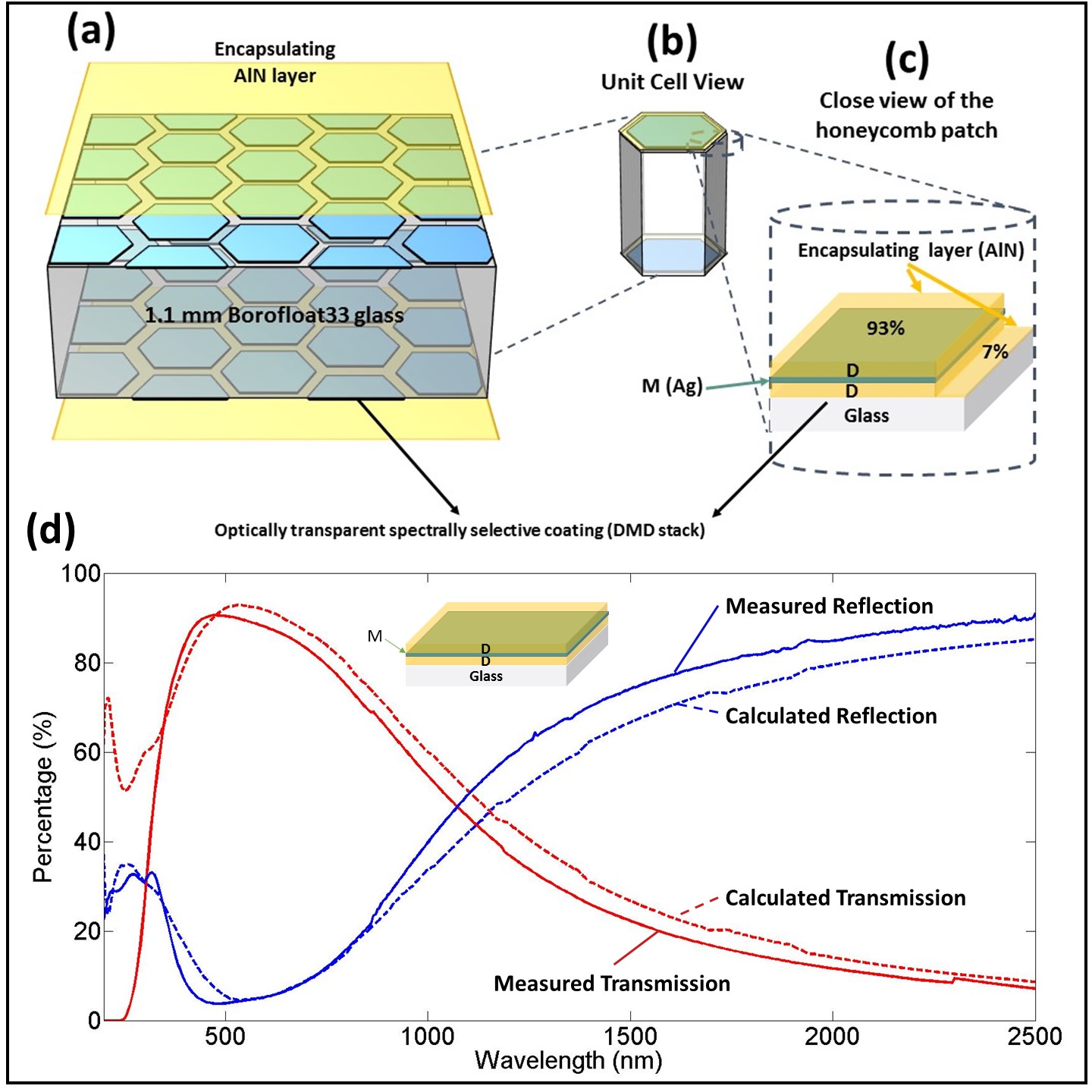}
    \caption{a) Meta-glass: a symmetrically layered structure made of identical metasurfaces on both faces of the glass substrate. (b) Unit cell view of the meta-glass comprising aligned honeycomb patch arrays on both sides. (c) Honeycomb patches made of a cascading dielectric-metal-dielectric (DMD) spectrally selective coating where 93\% is the fill fraction. (d) Comparison between the modelled data based on the transfer-matrix method and the measured data for the transmission and reflection of the uniformly coated glass substrate comprising the designed DMD on one side, wherein the respective layer thicknesses (from the bottom - up) are total 53 nm / 10 nm / 45 nm.
}
    \label{fig:f7}
\end{figure}

\subsection{DMD Optical Coating: Visible Transparency, NIR Control and Low Thermal Emissivity}
In order to design a highly electrically conductive transparent coating with high NIR reflection and low thermal emissivity, we employ a uniform dielectric-metal-dielectric (DMD) spectrally selective coating comprising AlN (i.e., D) and Ag (i.e., M) layers (see Fig, 1 c). We subsequently utilize the designed DMD coating to create the honeycomb patch patterned conductive metasurfaces on both sides of the glass substrate to enable RF transparency. This two-step design procedure first optimizes the uniform (unpatterned) DMD coating performance metrics in the visible, NIR and MIR, and thereafter utilizes this high electrical conductivity coating for the metamaterial design in the RF region.

Highly transparent metallo-dielectric coatings in the visible region are achieved by availing wave-interference effects through variation in the dielectric and metal layer thicknesses. We analyze the optical coating using the well-established transfer-matrix method where the transfer matrices describing each layer are used to calculate the overall spectral transmission and reflection \cite{katsidis2002general,macleod2017thin}. A detailed derivation of reflection and transmission coefficients for the multilayer structures (e.g., DMD) are given in the supporting information (SI). 

Following derivation of the equations describing the transmission and reflection coefficients (eqs (12) and (13) in  the SI), we use the genetic algorithm (GA) toolbox in MatLab to optimize the optical response. In order to obtain the optimal optical performance, high transparency in the visible and NIR control, we maximize a weighted transmission parameter (visible transmittance, $T_{vis}$) using the photopic sensitivity function ($V(\lambda)$), where the latter is the spectral response of an average human eye (Eqn. 1). Further, the average reflection coefficient of the DMD coating in the near-nfrared is maximized to achieve the desired NIR control along with high visible transparency \cite{ko2018ultrasmooth} (Eqn. 2). Nominal Ag thickness of 10 nm was selected to achieve high visible transmission and thermal emissivity of $\epsilon \approx 0.1$ for each uniform (unpatterned) DMD layer stack. We used the optimization constraint of $T_{vis}>80\%$ to ensure high visible transparency and thus obtained the optimal metal/dielectric layer thicknesses.
\begin{equation}
    T_{vis}=\int_{400 nm}^{700 nm} V(\lambda)t_{DMD-coating}(\lambda) d\lambda
\end{equation}
\begin{equation}
    R_{NIR}=\frac{1}{2500-700}\int_{700 nm}^{2500 nm} r_{DMD-coating}(\lambda) d\lambda
\end{equation}
$r_{DMD-coating}$ and $t_{DMD-coating}$ are the reflection and transmission coefficients of the DMD coating, respectively. The optimal DMD stack comprising AlN/Ag/AlN layers with respective thicknesses of  53 nm/10 nm/45 nm yields $R_{NIR}=66\% $ and $T_{vis}=83\% $. The interested reader is referred to Ko \textit{et al.}  \cite{ko2018ultrasmooth} for a detailed discussion of these spectrally-selective metallo-dielectric optical coatings.

Thermal emissivity of a glass window represents the mid-infrared or thermal energy radiated by the glass surface. Low thermal emissivity coatings on windows play the important function of markedly reducing the loss of thermal energy from buildings. As such, the integration of visibly transparent, low thermal emissivity coatings enhance the thermal insulation properties of windows and thus promote energy conservation. Thermal emissivity estimates can be obtained by recognizing that it is directly proportional to the electrical sheet resistance of a metallo-dielectric film. The Hagen-Rubens formula approximately describes the relationship between emissivity and sheet resistance\cite{ibach1999freie}.
\begin{equation}
    \epsilon=1-r\approx\frac{4R}{Z_0}
\end{equation}
where $r$ is the reflection coefficient in the MIR range, $R$ is the sheet resistance of the DMD film and $Z_0=377 \Omega$ is the intrinsic impedance of free space. The sheet resistance of spectrally selective DMD films as a function of the Ag thickness is presented in the SI. The emissivity of the DMD film comprising 10 nm Ag is estimated to be $\epsilon=0.068$.
We note that the significance of low thermal emissivity and thermal resistance of the multi-paned windows is well established in the literature \cite{sadooghi2018thermal}. Specifically,  Sadooghi and Kherani present the thermal properties of triple and quadruple pane windows as a function of the optical properties of each pane; each pane comprises of glass or equivalent substrate with a visibly transparent, NIR control and low-emissivity coating and the interpane gaps filled with gases of various thermal conductivities. Accordingly, the objective herein is to create a meta-glass structure where the optical and thermal properties are substantially preserved, and thus its energy conserving function, yet integrating the desired RF transmission.   


Figure 1d illustrates that the modelled $T_{vis}^{DMD}$ compared to the measured data for the uniform (unpatterned) coating are in good agreement. Next, we present the design of a composite metasurface comprising two identical metasurfaces on each side of a glass substrate; and in particular, the metasurface here is a  honeycomb patch array patterned out of the designed metallo-dielectric coating.

\subsection{RF transparent Honeycomb-based Metasurface}

We now present the design of an RF transparent composite meta-glass structure operating at 30 GHz. Considering the highest degree of two-dimensional symmetry, the nature-inspired honeycomb patch array metasurfaces are utilized as two parallel surface impedances that mimic the performance of a double-stub impedance matching transmission line (i.e., double-stub analogy) in free-space (see Fig. 1 and 2)\cite{pozar2009microwave}. Given both angular and rotational symmetry, the proposed structure possesses high polarization fidelity along with a wideband, wide-angle, RF response which is highly relevant for radar applications \cite{he2019thin,Safari2020optically}. 

\subsection{Double-stub Impedance Matching}
Transmission-line theory has been extensively used to design matching circuit elements \cite{papapolymerou2003reconfigurable}. Double-stub impedance matching is a popular technique in the field of microwave engineering to reduce the reflection from a load in a transmission line \cite{pozar2009microwave}. This approach normally uses two shunt stubs (i.e., short/open lines parallel to the transmission line) as parallel inductive/capacitive tunable elements to achieve impedance matching by changing the length of each stub. Here, we analogously use this impedance matching method to achieve RF transparency in free-space. We begin by considering two identical surface impedances on both sides of the glass substrate (see Fig. 2) and examine the conditions for impedance matching and accordingly we then construct the honeycomb patch metasurfaces. The transmission matrix of the proposed structure is obtained in the following way. 
\begin{equation}
    M_{meta-glass}=M_g M_{glass} M_g
\end{equation}
\begin{equation}
   M_g= \left(\begin{matrix}
    1 & 0 \\
    jB_{honeycomb\ patch\ array} & 1
\end{matrix}\right)
\end{equation}
\begin{equation}
     M_{glass}= \left( \begin{matrix}
    \cos k_{glass}L_{glass} & iZ_{glass}\sin k_{glass}L_{glass} \\
    \frac{i}{Z_{glass}}\sin k_{glass}L_{glass} & \cos k_{glass}L_{glass}
    \end{matrix}\right)
\end{equation}
where, the subscripts $g$ and $glass$  denote the matrix parameters associated with the impedance of the array and the glass substrate, respectively. $B$ represents the susceptance of each impedance surface, which can be realized using the capacitive honeycomb-patch array and tuned by virtue of changing the geometrical parameters. The reflection and transmission coefficients of the proposed double-stub matching structure can be calculated using equations (12) and (13) described in the SI. Figure 2 shows the transmission value with respect to the susceptance of the surface grid (i.e., $B_{honeycomb\ patch\ array}$). It is clear from this figure, that the unity transmission can be achieved with both inductive (i.e., $B_{honeycomb\ patch\ array}= -0.0056 \ S$) and capacitive (i.e., $B_{honeycomb\ patch\ array}= +0.0056 \ S$) surface impedances. 
\begin{figure}
    \centering
    \includegraphics[width=15cm]{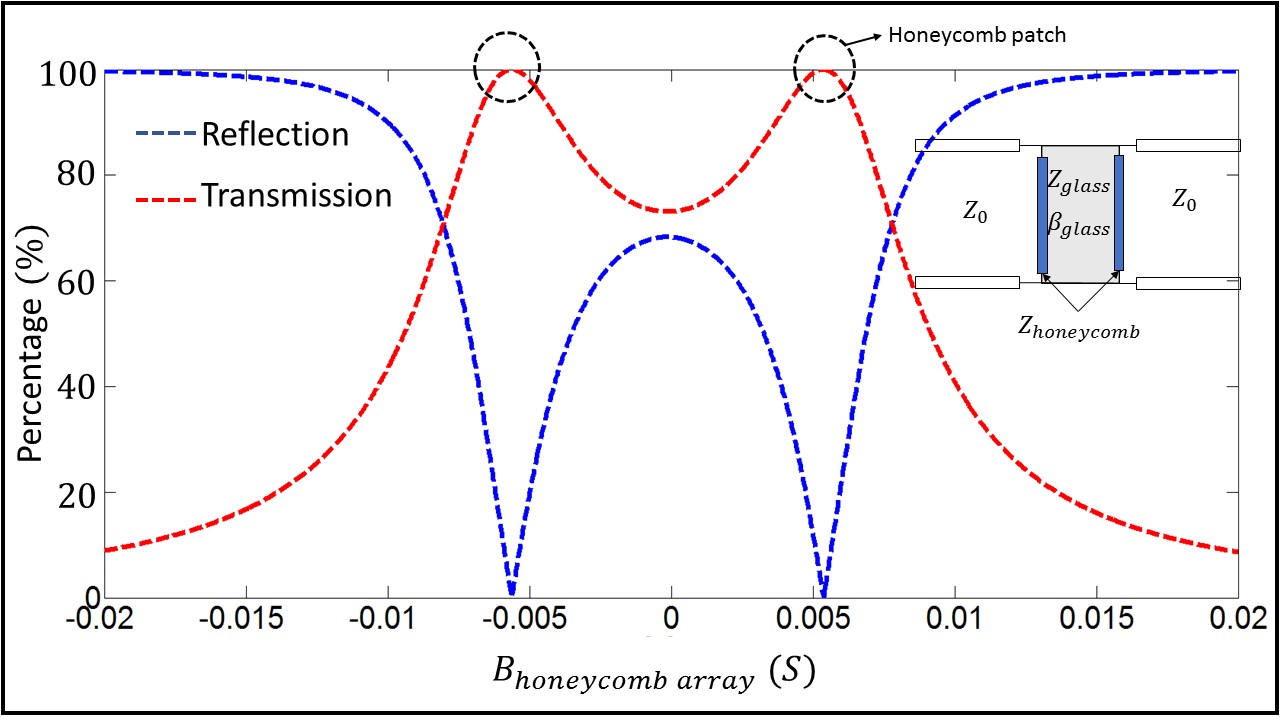}
    \caption{Double-stub impedance matching: RF frequency response as a function of the surface susceptance (i.e., $B_{honeycomb}$ $(S)$, $Y_{honeycomb}=G_{honeycomb}+jB_{honeycomb}$) of the honeycomb metasurface.}
    \label{fig:f7}
\end{figure}
Having shown the viability of impedance matching using two surface impedances, next using the honeycomb patch array we determine the relevant parameters that yield the desired surface impedance (i.e., $B_{honeycomb\ patch\ array}= +0.0056 \ S$) and thus obtain unity transmission at 30 GHz for the composite metasurface. 

\subsection{Honeycomb-based Composite Meta-glass}

In the supporting information, we present the ABCD transfer-matrix method and derive the surface impedance for a single honeycomb patch metasurface on glass. Here, we employ the transfer-matrix method to calculate the transmission and reflection coefficients of the composite meta-glass structure comprising two honeycomb patch metasurfaces on both sides of glass substrate, following the double-stub impedance matching analogy (Fig. 3a), in which the transmission and reflection of the composite honeycomb-based meta-glass is calculated using equations (12), (13), and (14) (see SI). It is clear from Fig. 3b that the calculated reflection coefficient is in good agreement with the numerical simulations.

\begin{figure}
    \centering
    \includegraphics[width=15cm]{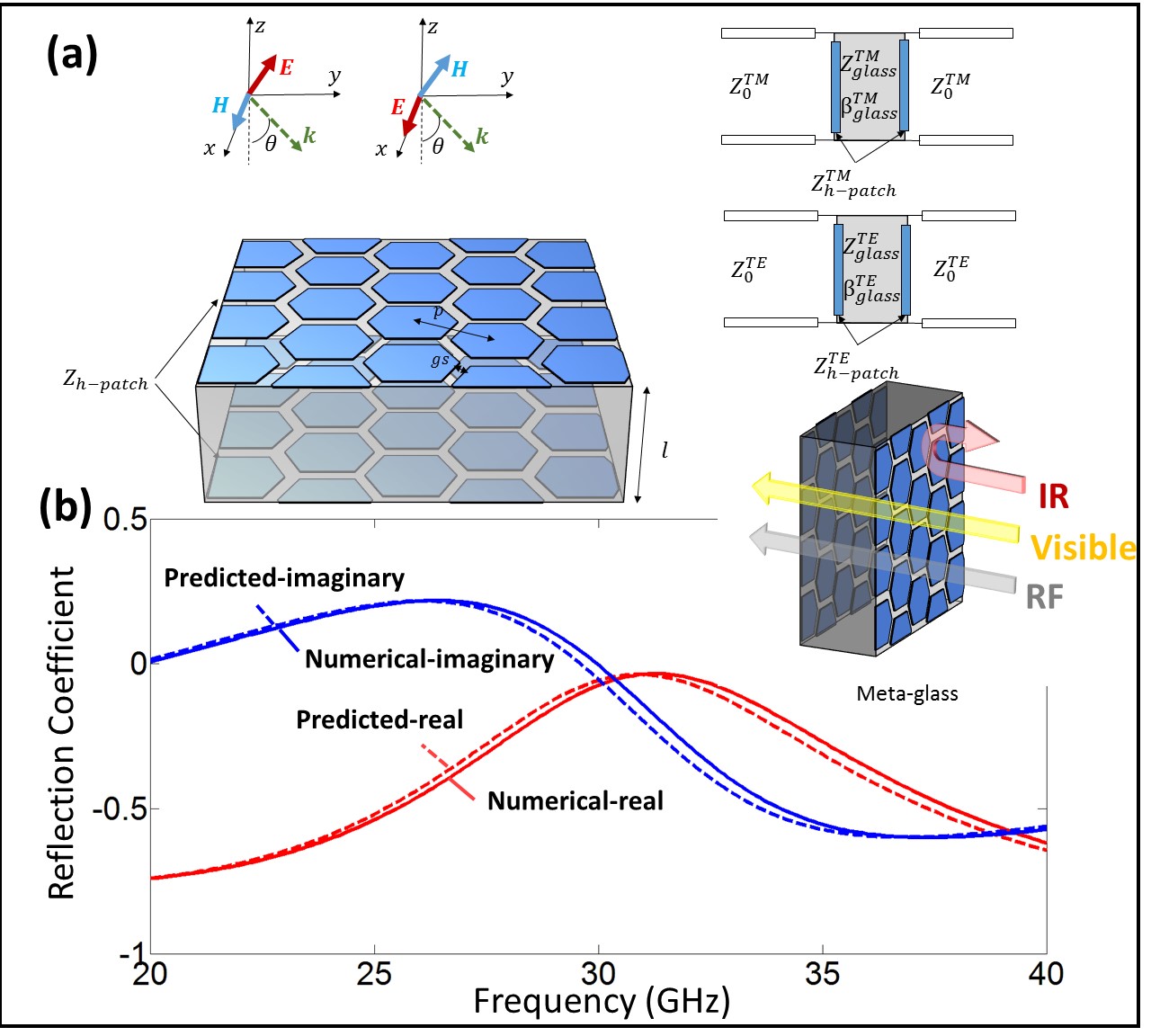}
    \caption{(a) A 3D schematic of the proposed meta-glass structure ($p=800 \mu m$, $gs=50 \mu m$, and $l =1.1 mm$) and (b) the validated RF reflection profiles determined using the numerical and analytical (predicted) modeling techniques.}
    \label{fig:f7}
\end{figure}
The frequency response of the matched structure, $B_{honeycomb\ patch\ array}=0.0056 \  S$, comprising a glass substrate of 1.1 mm thickness and relative permittivity of 5.3 is derived and shown in Fig. 3 b and 4. Both figures show good agreement for the angular and frequency responses determined analytically and through numerical modelling, validating the assumption of modelling each metasurface as an admittance surface on the glass substrate. 
The angular and frequency dependent responses of the designed meta-glass structure are illustrated in Fig. 4. We observe that the designed composite metasurface possesses wideband and wide-angle RF transparency centered at 30 GHz. In the following sections we present experimental data corroborating the optical and RF performance of the designed composite meta-glass. 
\begin{figure}
    \centering
    \includegraphics[width=15cm]{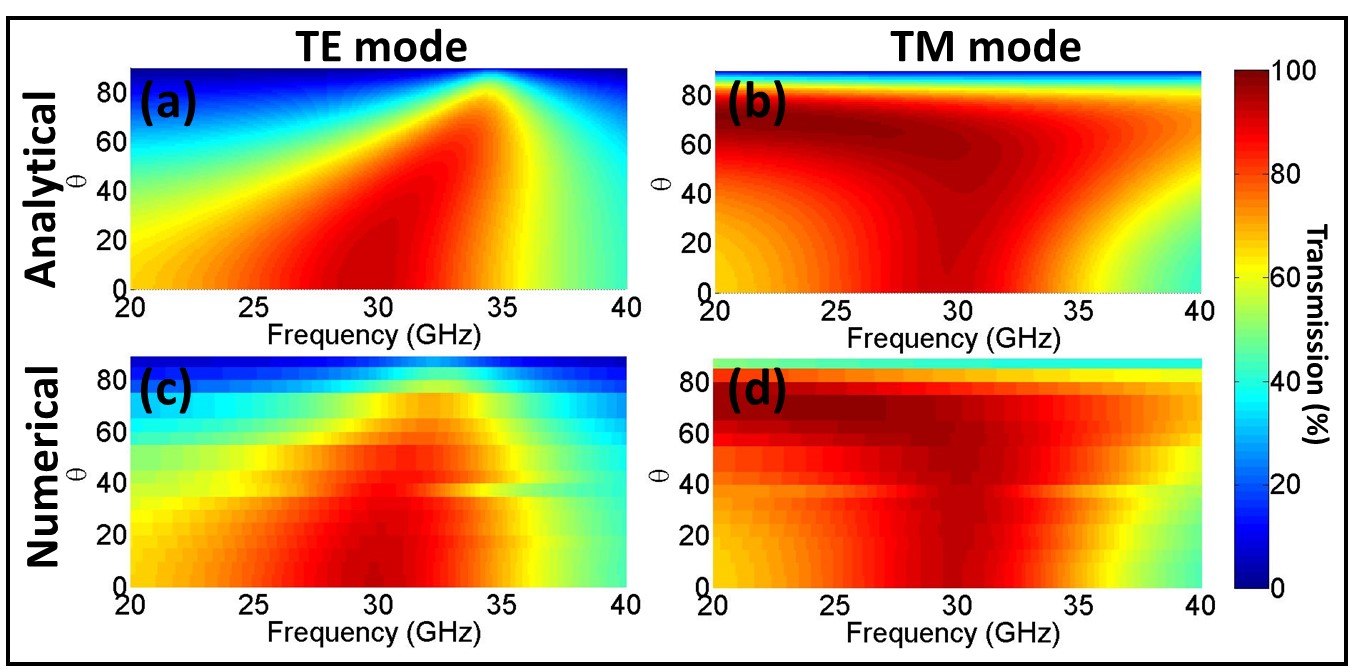}
    \caption{Analytically modelled angular RF transmission under (a) TE- and (b) TM-polarized incident waves, and the numerically modelled RF response under (c) TE- and (d) TM-polarized incident waves for the meta-glass structure.}
    \label{fig:f7}
\end{figure}

While the proposed transfer-matrix method is capable of designing multilayer composite metasurfaces with large substrate thicknesses, alternatively the effective parameter retrieval method can be used to analyze and design relatively small but finite (compared to the operating wavelength) impedance matching structures. In other words, the effective permittivity and permeability of the entire structure can be engineered to match that of free-space and achieve unity impedance. We refer the interested reader to a prior study which presents the underlying theory on impedance matching of a meta-glass comprising of honeycomb mesh grids on both sides of a glass substrate \cite{Safari2020optically}.

\subsection{Flexible Design: Amenable to Different Substrate Thicknesses and Operating Frequencies}
Wireless telecommunication frequencies have drastically changed over the past decade. High-frequency communication helps increasing the available bandwidth (i.e. the speed of data transfer). However, due to higher material and free-space losses in the smart city environment, the frequency range is ultimately limited by the efficiency of the communication technology and health standards. Moreover, the advent of advanced communication technologies (4G and 5G) call for a versatile and flexible design that can account for continual increase in operating frequencies of newly developed communication devices. We show that the designed composite meta-glass structure is amenable to higher frequency operation as a result of frequency scalability of the metasurface design.  

\begin{figure}
    \centering
    \includegraphics[width=15cm]{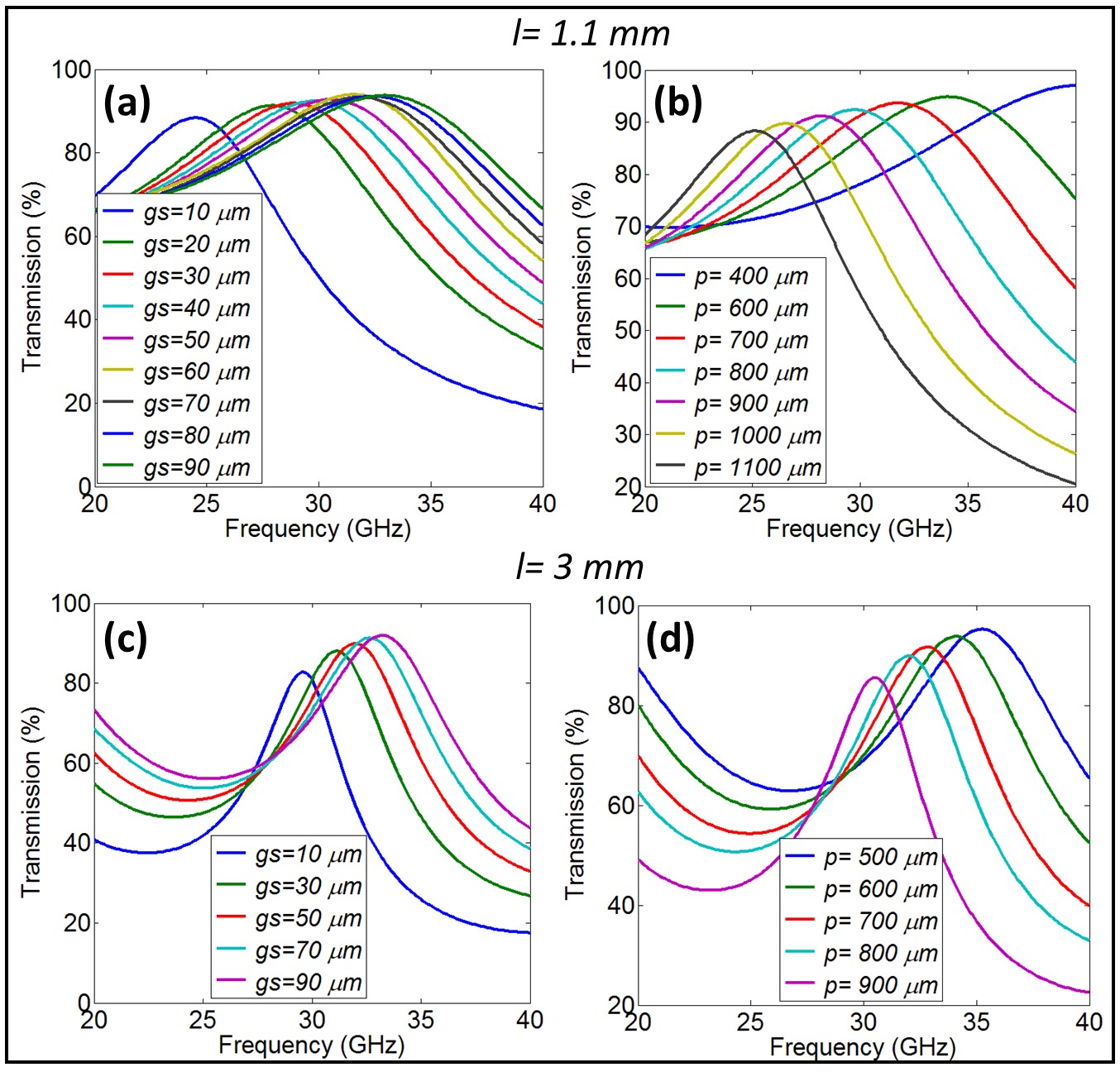}
    \caption{Transmission of the meta-glass for honeycomb patch arrays with different (a) gap sizes (i.e., $gs$) and (b) periodicity values (i.e., $p$)}
    \label{fig:f7}
\end{figure}

The honeycomb patch array can be engineered to achieve RF transparency for a certain substrate thickness and at a particular operating frequency. The impedance of the honeycomb patch array can be tuned by virtue of changing the geometrical parameters, that is engineering the periodicity $p$ and gap size $gs$. Since a fixed gap size ($gs$) of 50 $\mu m$ is deemed to minimize optical contrast from the perspective of the average human eye, RF frequency response of the meta-glass structure for  different periodicity ($p$) values and fixed $gs$ are shown in Figs. 5b and 5d. It is worth noting that both the periodicity and gap size affect the efficiency of RF transparency, the operating frequency, and the bandwidth.

Figure 5 illustrates the RF frequency response for two substrate thicknesses ($l_{glass}$) of 1.1 $mm$ and 3 $mm$. The semi-analytical design of the meta-glass based on the double-stub method allows us to take into account the substrate thickness and accordingly derive the required surface impedance (see Eqn. (6)). The input impedance of the double-stub-based impedance matching structure (i.e., the meta-glass) can be tuned to match the intrinsic impedance of the glass substrate with that of the surrounding medium. 

We now present the experimental results describing and verifying the performance of the designed meta-glass structure in the sections to follow. 

\section{Experimental Results and Discussion}
The proposed meta-glass is a versatile design capable of wideband and wide-angle impedance matching at 30 GHz which is visibly transparent and low-emissive in the IR range. The meta-glass is fabricated by utilizing the optimized DMD optical coating comprising Ag and AlN layers, that are patterned on both sides of the glass substrate. 

As it was mentioned in the introduction, the authors discussed the main properties of a honeycomb mesh meta-glass design, which is visibly and RF transparent \cite{Safari2020optically}. The honeycomb mesh can potentially exploit the interconnected silver meshes incorporated in its design for active defogging through electrification of the conductive mesh. However, this design lacks high IR-control and low-emissivity which is of high interest in energy conservation applications. Also, due to the large size of the hexagonal patterns, the optical contrast makes the patterns slightly visible to the naked human eye. With the understanding of these limitations, the herein proposed honeycomb patch arrays are designed to make use of a high silver surface coverage ratio on both sides of the glass substrate and obtain high IR-control and low-emissivity. Moreover, due to the small gap size between the hexagonal patches, the contrast is not visible to the average human eye and the structure seems to be more transparent. Due to the presence of narrow traces forming the metallic meshes, higher surface current densities are observed, which gives rise to higher ohmic losses and lower RF transparency of the honeycomb mesh meta-glass\cite{Safari2020optically} when compared to the patch meta-glass structure. As a result, the honeycomb patch meta-glass outperforms the honeycomb mesh structure under different incident angles. Therefore, the proposed honeycomb patch array meta-glass is a superior alternative to the previous design both in RF and optical region. Also, it is worth noting that due to lower ohmic losses, a honeycomb patch array composed of a  thinner silver layer is used which also makes this structure a superior design economically.
Hereunder, we experimentally explore and study the RF and optical performance of the proposed meta-glass structure.

\begin{figure}
    \centering
    \includegraphics[width=13cm]{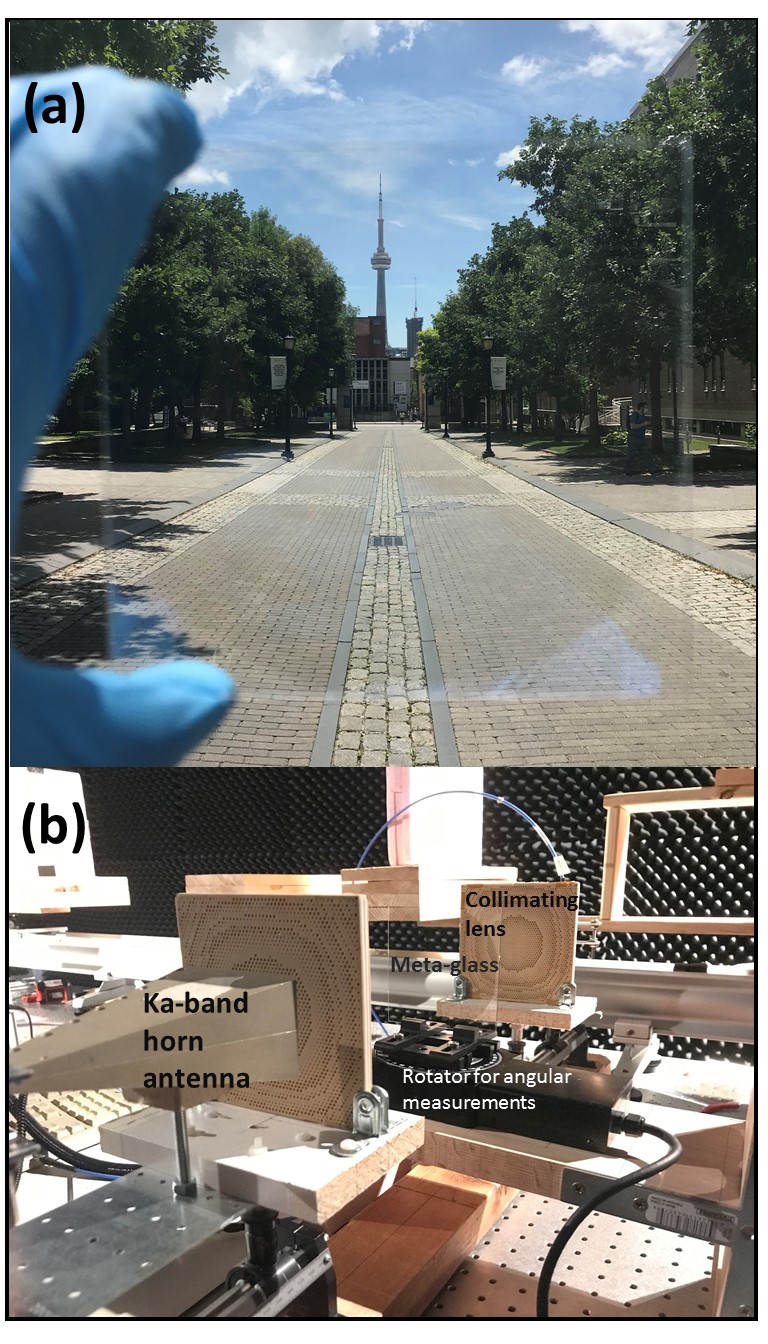}
    \caption{(a) The fabricated meta-glass structure illustrating its visual clarity and (a) the RF measurement apparatus.}
    \label{fig:f7}
\end{figure}
\subsection{Optical Performance}
A detailed explanation of the fabrication process for the metallo-dielectric spectrally selective coating is given in the literature\cite{ko2018ultrasmooth,Safari2020optically}. A stepwise description of the process of patterning the DMD optical coating using photolithography and lift-off so as to realize the honeycomb patch array is given in\cite{Safari2020optically}. We now present experimental measurements carried out on the fabricated structure (Fig. 6a);  the measurement procedure and apparatus are described in the Appendix. Figure 7 compares the transmission and reflection profiles of the uniform (unpatterned) low thermal emissivity DMD coating on glass (low-emissivity glass) with that of the bare-glass substrate and the proposed meta-glass structure in the visible and NIR regions. It is evident that the NIR reflection of the meta-glass is slightly lower than that of the low-emissivity glass while being 75\% higher than that of the plain glass. These spectral profiles illustrate that the proposed meta-glass preserves the optical properties of the low-emissivity glass by supporting high transparency in the visible and NIR reflection while possessing high RF transparency.

\begin{figure}
    \centering
    \includegraphics[width=15cm]{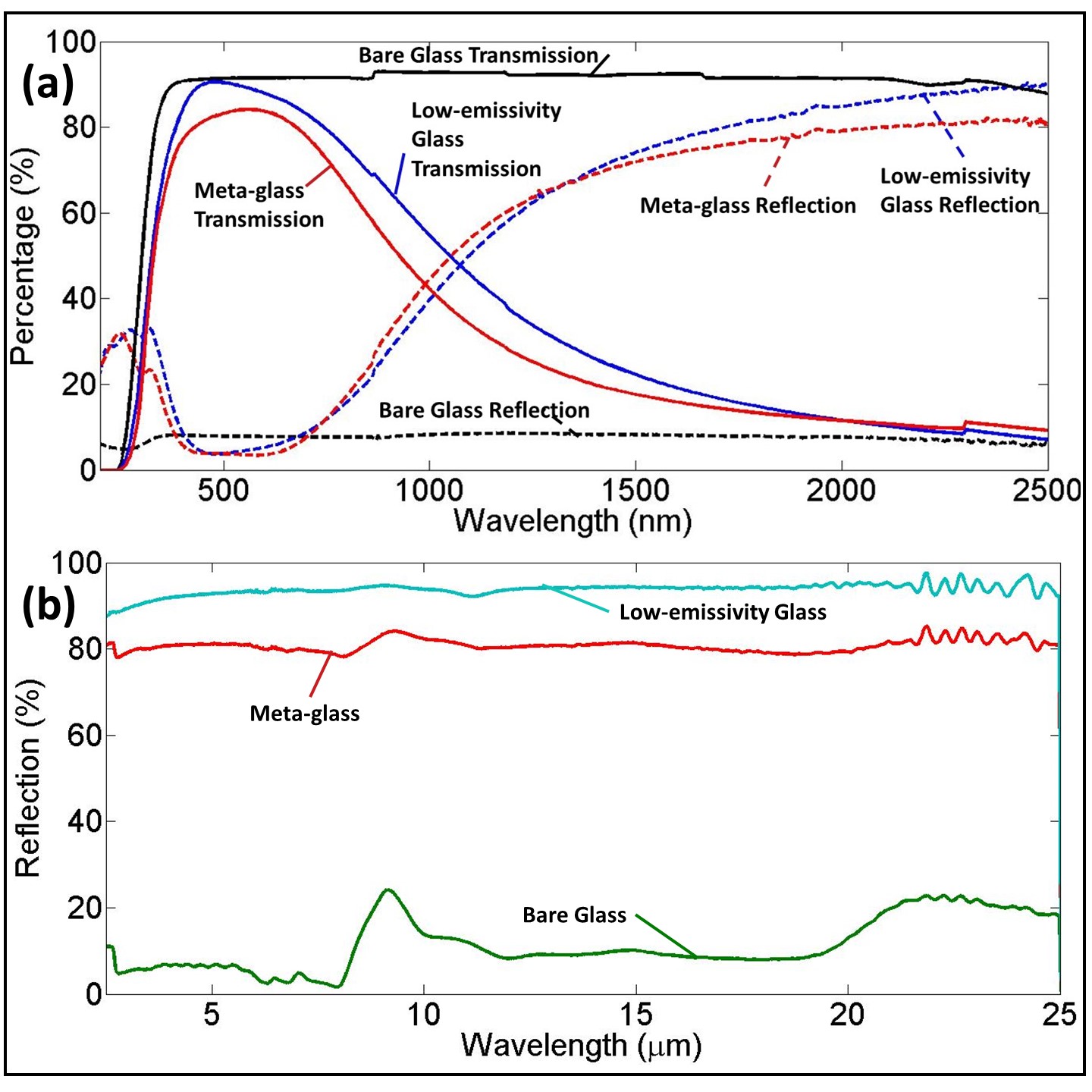}
    \caption{a) Spectral optical response of the meta-glass, bare-glass (both sides uncoated), and the low-emissivity glass (one-side uniformly coated (unpatterned)) b) FTIR measurements using the specular reflection accessory, where the incident angle of illumination is 30$^{o}$.}
    \label{fig:f7}
\end{figure}
Peak visible transparency of $83 \%$ is measured at $\lambda = 550 nm$ and the $>60\%$ NIR reflection is measured at $2500 nm>\lambda>1000 nm$ range. Moreover, using an FTIR spectrometer (see Appendix),  MIR reflection of the meta-glass is determined to be $>80 \%$ in the $2.5 \mu m<\lambda<25 \mu m$ range; in contrast the MIR reflection of the low-emissivity glass is $92 \%$  over the same wavelength range (both measurements having been carried out under illumination at an angle of incidence of 30$^o$ ). Emissivity of the uniform (unpatterned) film was estimated to be $\epsilon=0.068$ which indicates a reflection ($r \approx 1-\epsilon$) of $93\%$ , which is in good agreement with the measured MIR reflection of $92\%$. The lower emissivity of the meta-glass compared to that of the low-emissivity glass is due to the lower surface coverage of silver containing metallo-dielectric coating commensurate with metasurface patterning. The emissivity of the meta-glass could be enhanced by increasing the thickness of the Ag layer albeit the metallo-dielectric coating would require appropriate optical tuning to achieve the desired transmissivity in the visible and reflectivity in the NIR. Even though the emissivity of the meta-glass is slightly higher than that of the low-emissivity coating, the measurement shows 70 \% enhancement in MIR reflection when compared to the plain glass substrate.

We note that while the foregoing design and demonstration are based on a single metallic layer within a three-layered (DMD) metallo-dielectric spectrally selective coating, it is possible to extend the basic metallo-dielectric coating design to include multiple metallic layers whereby the sheet resistance can be markedly reduced (increase in electrical conductance) and accordingly increasing the NIR reflection and reducing the thermal emissivity (increase in MIR reflection). With multiple metallic layers the optical transparency can be maximized by appropriate tuning of the dielectric and metallic layers. These advanced coatings with multiple metallic layers then become the basis of the composite metasurface designs. The theory and design of multiple metallic layer based metallo-dielectric coatings is described by Safari \textit{et al.}\cite{Safari2020optically}. 

\subsection{RF Performance}
We now present experimental validation of the RF design where the measurements and associated analysis are based on a quasi-optical characterization apparatus (Fig. 6 b). The RF measurement technique is explained in detail in the Appendix. Spectral RF transmission and reflection coefficients of the composite meta-glass structure, bare glass, and the low-emissivity glass are shown in Fig. 8. It is clearly observed that the RF transmission of the meta-glass structure outperforms that of the bare-glass and the low-emissivity glass. In contrast to the low-emissivity glass, which blocks the RF signal entirely, the meta-glass allows $92\%$ of RF transmission and it is also substantially greater than the $70\%$ RF transmission of the glass substrate. This high transmission is attributed to the two identical optically transparent honeycomb-based surface impedance layers on each side of the glass substrate which are optimized using the double-stub impedance matching technique. The effective permittivity and permeability of the composite metasurface is near unity at normal incidence and accordingly yields wide-angle impedance matching (see Fig. 4).
\begin{figure}
    \centering
    \includegraphics[width=15cm]{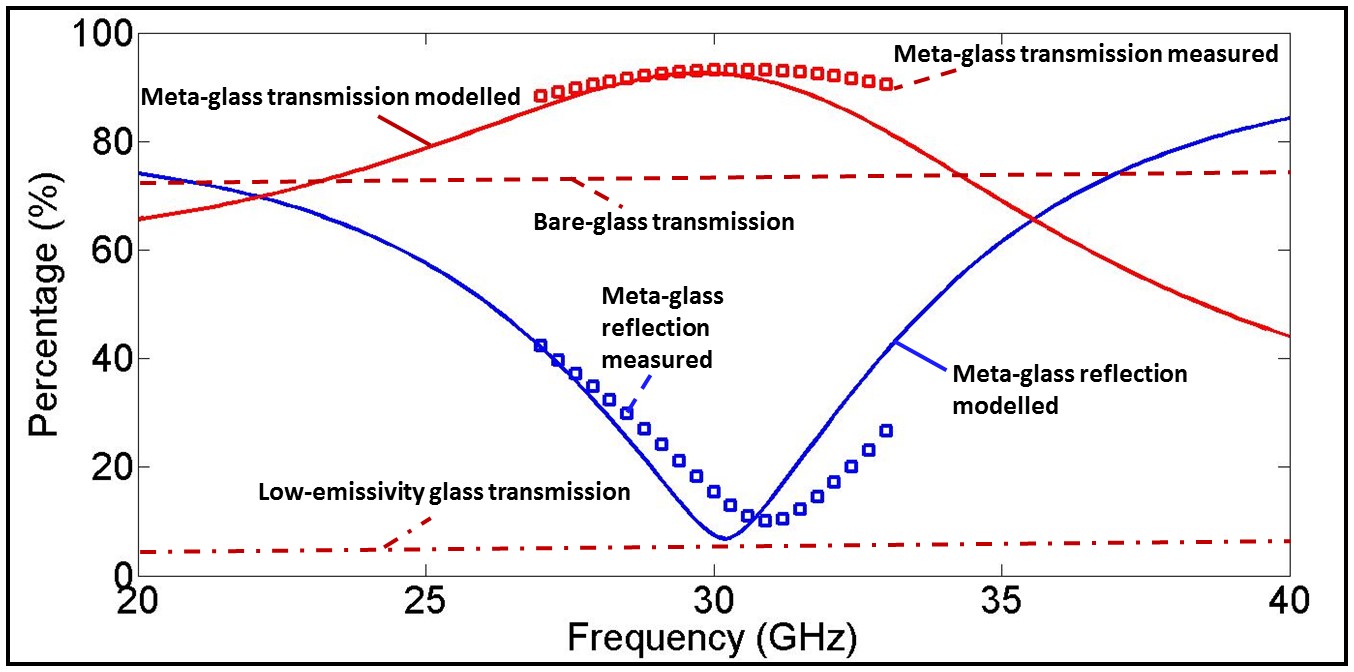}
    \caption{RF transmission and reflection coefficients for the meta-glass (modelled and measured), bare-glass (measured), and low-emissivity glass (measured)}
    \label{fig:f7}
\end{figure}

The meta-glass outperforms the plain glass in the 25-35 GHz range, the operating bandwidth of the composite metasurface. The high 2-dimensional symmetry of the honeycomb patch array results in high polarization fidelity which in turn sustains high RF transmission under small variations in  incident angle and polarization (see the SI). The confluence of all the properties in the visible, NIR, MIR and RF frequency ranges make the meta-glass amenable to a manifold of window applications which include  automobile radar, radoms, and low-emissivity solar control windows.

\section{Conclusion}
In conclusion, we have shown that a composite metasurface comprising of patterned spectrally selective dielectric-metal-dielectric (DMD) coatings can simultaneously provide high transparency in the visible and RF regions, high NIR reflection and low thermal emissivity. Theoretical analysis and experimental corroboration are presented to demonstrate the multi-functional performance of the designed meta-glass over a range of frequencies. The nature-inspired honeycomb patch array is particularly apropos given its inherent 2-dimensional symmetry and thus high polarization fidelity within the composite metasurface design. Intrinsic frequency scalability of metamaterials is availed to achieve RF transparency over a variety of RF frequency bands; that is, frequency tuning is possible by varying the geometrical parameters of the honeycomb patch array (gap size $gs$ and periodicity $p$). In particular, we demonstrate a meta-glass having $gs=50 \mu m$ and $p=800 \mu m$ capable of operating at 30 GHz with a 10 GHz bandwidth. We obtain 92\% peak RF transmission and 83\% visible transparency at 30 GHz and 550 nm, respectively. The meta-glass exhibits $>$ 60\% near-IR reflection and $>$ 80\% mid-IR reflection which is equivalent to low thermal emissivity of $\epsilon \approx0.2$. These properties make the meta-glass a viable candidate for energy conservation and 5G communication applications and a 5G-compatible substitute for  existing low-emissivity glass. This comprehensive proof-of-concept paves the way for future advances in the development of transparent and tunable metamaterials for 5G and 6G communication, space applications, metamaterial computational processors and beyond.

\section{Appendix: Fabrication}
We used photolithography, RF sputter deposition of AlN/Ag/AlN (DMD) spectrally selective thin films, and lift-off techniques to fabricate the proposed meta-glass structure. The sputter deposition technique for metallo-dielectric stacks is explained in detail in the literature\cite{ko2018ultrasmooth}. Metal lift-off and photolithography processes were used to identically fabricate the patterns on both sides of the glass substrate. Figure 9 illustrates the fabrication steps sequentially: we begin with 10 cm $\times$ 10 cm plain glass (1.1 mm thick Corning Eagle XG glass) as the substrate. UV light source - mask aligner was used to expose honeycomb patch patterns on the S1818 photoresist which was spin-coated on both sides of the glass. The designed spectrally selective DMD coatings (each coating on each glass surface having  a Ag thickness of 10 nm and thus the total Ag thickness of 20 nm in the meta-glass) were deposited on the patterned photoresist on both sides of the glass substrate. Finally, lift-off technique was used to develop the honeycomb-shaped pattern on each side of the glass substrate and thus the designed hexagonal meta-glass structure.
\begin{figure}
    \centering
    \includegraphics[width=15cm]{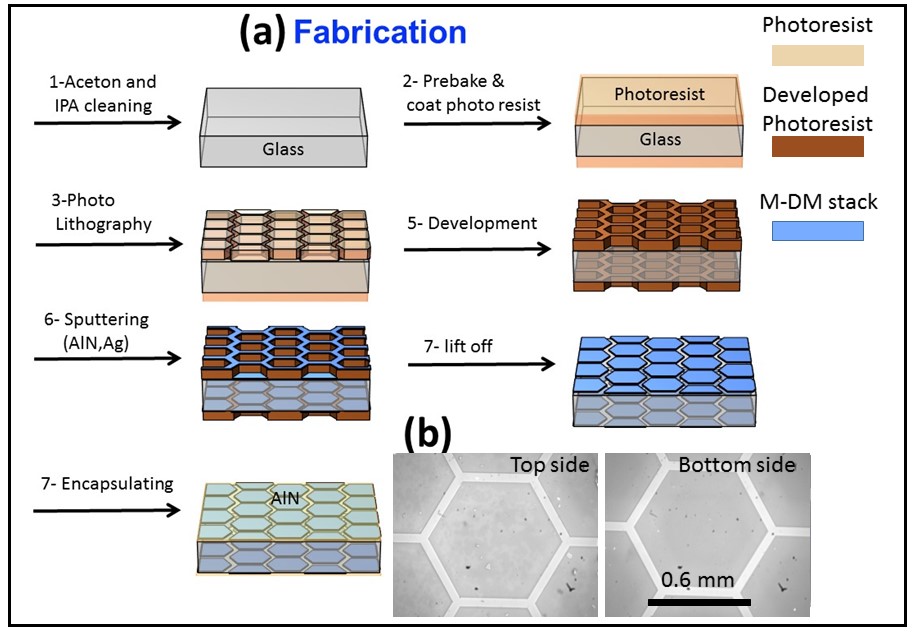}
    \caption{(a) A step-by-step demonstration of the fabrication process and (b) microscopic images of the honeycomb patch array.}
    \label{fig:f7}
\end{figure}
\subsection{RF Measurements} 
The quasi-optical microwave apparatus consisted of a Vector Network Analyzer (VNA), two k-band horn antennas, and confocal lenses optimized to operate at (25 – 40) GHz (Fig. 6 b). The Gaussian beam-waist (beam width of 4 cm) is located at the center, where the sample is positioned, and is designed to cover 10 hexagonal unit cells, and is small enough so as to avoid any spillover of the incident wave on the 10 cm$\times$10 cm sample \cite{he2019thin}(Fig. 6 b).
\subsection{Optical Measurements} 
Optical transmission and reflection measurements were carried out in the UV, visible, and IR regions (200-2500 nm) using the PerkinElmer Lamda 1050 UV/VIS/NIR spectrometer.
FTIR measurements were carried out using the Pike Technologies VeeMax II variable angle specular reflectance accessory using Perkin Elmer Spectrum Two. FTIR reflection measurements were obtained over the spectral range of 2.5-25 $\mu m$ under illumination at 30$^{o}$  from the normal.
\begin{acknowledgement}
The authors gratefully acknowledge the assistance of Min Kim in Metawaves group for fruitful discussions on the RF measurement apparatus and Remy Ko and Rajiv Prinja in the Advanced Photovoltaics-Photonics and Devices (AP$^2$D) Labs with regard to key discussions on the fabrication of the patterned transparent conductive coatings. The support of Drs. Navid Soheilnia, Laleh Mokhtarpour and Anantha Jain, of 3E Nano Inc., in the design, fabrication and characterization of the spectrally selective coatings is gratefully recognized. Financial support from the Natural Sciences and Engineering Research Council of Canada (NSERC)  through a Strategic Project Grant, MTI Technologies, 3E Nano Inc., CMC Microsystems, and the University of Toronto is gratefully acknowledged.

\end{acknowledgement}


\begin{suppinfo}

\section{Transfer Matrix Method}
We analyze the optical coating using the well-established transfer-matrix method relating the electric and magnetic fields at the beginning (e.g., $E(z)$) and the end (e.g,, $E(z+L)$) of each layer and thus calculating the overall transmission and reflection\cite{katsidis2002general,macleod2017thin}.
\begin{equation}
     \left( \begin{matrix}
    E(z+L) \\ 
    H(z+L)
    \end{matrix}\right)
    = M \left( \begin{matrix}
    E(z) \\ 
    H(z)
    \end{matrix}\right)
\end{equation}
    \begin{equation}
    M= M_{d}.M_{m}.M_{d}=
    \left(
    \begin{matrix}
    A & B \\ 
    C & D 
    \end{matrix}
    \right)
    \end{equation}
\begin{equation}
    M_{m/d/glass}= \left( \begin{matrix}
    \cos \delta_{m/d/glass} & i\gamma_{m/d/glass}\sin \delta_{m/d/glass} \\
    \frac{i}{\gamma_{m/d/glass}}\sin \delta_{m/d/glass} & \cos \delta_{m/d/glass}
    \end{matrix}\right)
\end{equation}
\begin{equation}
\delta_{m/d/glass}= k_{m/d/glass}L_{m/d/glass}\cos{\theta_{m/d/glass}}    
\end{equation}
\begin{equation}
\gamma^{TE}_{m/d/glass}= \frac{\eta_{m/d/glass}}{\cos{\theta_{m/d/glass}}},\  \gamma^{TM}_{m/d/glass}= {\eta_{m/d/glass}}{\cos{\theta_{m/d/glass}}}    
\end{equation}

where, $k$, $\eta$, and $\theta$ are the propagation constant, intrinsic impedance, and propagation angle of the wave inside each layer, respectively. Subscripts $m$, $d$, and $g$  denote the metal (i.e., Ag) layer, dielectric (i.e., AlN) layer, and the glass substrate, respectively, and $L$ is the respective thickness of each layer. Total transmission and reflection are obtained using the following equations at normal incidence \cite{katsidis2002general,macleod2017thin}.
\begin{equation}
    r=\frac{A+B/Z^{TE/TM}_0-CZ^{TE/TM}_0-D}{A+B/Z^{TE/TM}_0+CZ^{TE/TM}_0+D} 
\end{equation}
\begin{equation}
    t=\frac{2(AD-BC)}{A+B/Z^{TE/TM}_0+CZ^{TE/TM}_0+D}
\end{equation}
\begin{equation}
    Z^{TE}_0=\frac{Z_0}{\cos{\theta_i}}, Z^{TM}_0={Z_0}\cos{\theta_i}
\end{equation}
where $Z_0$ and $\theta_i$ are the intrinsic impedance of vacuum and the incident angle, respectively. $r$ and $t$ are the reflection and transmission coefficients, respectively, of the overall DMD coating.

\section{Thin Transparent DMD Film Conductivity}

The electrical conductance of a spectrally selective multilayer dielectric-metal-dielectric (DMD) stack plays an important role in the design of the meta-glass structure. We used a 4-point probe measurement technique to measure the sheet resistance of the DMD stack and thus the electrical conductivity. We utilized the obtained value to simulate the honeycomb patch metasurface using the thin film boundary condition in ANSYS HFSS. Moreover, in order to verify the validity of our simulations, we compared the numerically calculated S-parameter and the measured data using the VNA and quasi-optical measurement apparatus (data is presented in the main paper). It is observed in Fig. 10 that by increasing the comprising Ag thickness we can achieve higher conductivity, which in turn helps us achieve RF elements with lower insertion losses. While the thicker Ag layers help enhance the RF performance, the visible transparency of the DMD stack is generally reduced. To mitigate this drawback, multilayer dielectric-metal (m-DM) stacks is a viable path toward enhancing electrical conductance while maximizing visible transparency\cite{ko2018ultrasmooth} \cite{Safari2020optically}.

\begin{figure}
    \centering
    \includegraphics[width=15cm]{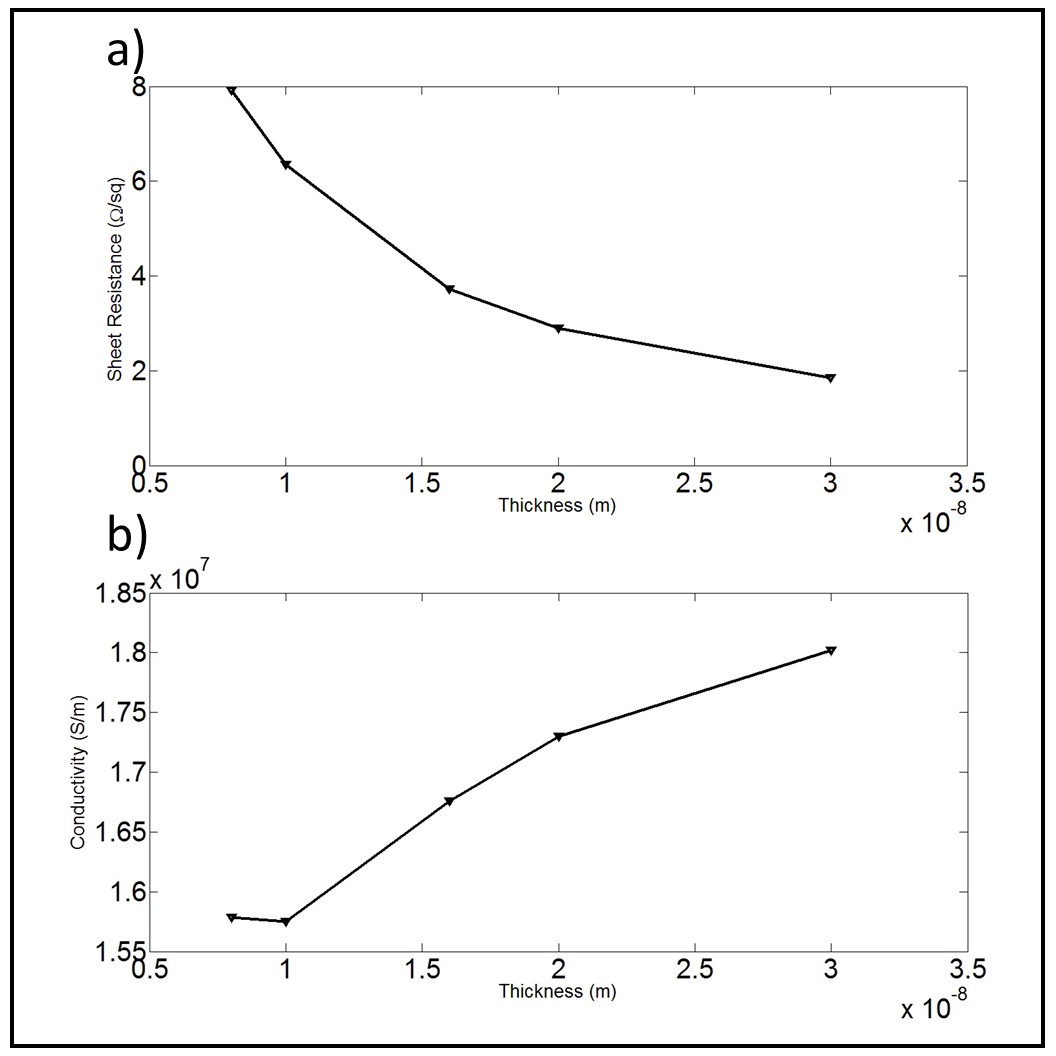}
    \caption{Sheet resistance (a) and bulk conductivity (b) of the thin film DMD stack as a function of the Ag thickness\cite{Safari2020optically}.}
    \label{fig:f5}
\end{figure}

\subsection{Honeycomb Patch Array}

In this section, we show that it is possible to use the honeycomb patch array metasurface as a surface impedance layer to achieve impedance matching and hence RF transparency. The derivation of surface impedance properties for square meshes and patches at arbitrary angles has been reported previously \cite{luukkonen2008simple,holloway2005reflection}. Even though the induced current distribution on hexagonal patches is different from that of square patches, through similar numerical simulations we calculate the surface impedance of the honeycomb patches. 


First, we consider an array of hexagonal (i.e., honeycomb) patches with gap size smaller than the periodicity of the array  $gs << p$, where $gs$ and $p$ are the gap size and periodicity, respectively. For the case of low frequency incident waves, that is, where the characteristic lengths such as the periodicity and gap size are much smaller than the operating wavelength, the structure can be considered isotropic and its electromagnetic response is weakly dependent on the choice of the incident plane. 

To characterize the composite honeycomb patch metasurfaces and tune the surface impedance, we consider a honeycomb patch metasurface patterned only on one side of a glass substrate (i.e., 1.1 mm thick Corning Eagle XG glass) (see Fig. 11). Here, the impedance matrix associated with the honeycomb patch metasurface is retrieved from the reflection and transmission coefficients at normal incidence. The geometrical parameters of the honeycomb patch array (i.e., periodicity $p$ and gap size $gs$) are tuned to obtain the desired susceptance $B_{honeycomb\ patch\ array}= +0.0056 \ S$, that is the susceptance value derived above using the double-stub impedance matching analogy technique to achieve unity transmission. In order to achieve low optical contrast, we take advantage of the human eye's limitation, which is its inability in detecting objects much smaller than 50 $\mu m$. Therefore, we set the gap size for the optimized structure at 50 $\mu m$. To achieve near-unity transmission at 30 GHz, a honeycomb patch array is designed with the periodicity of $p=800 \mu m$. The angular dependent RF transmission and reflection coefficients for the single honeycomb patch array on the glass substrate is calculated and verified using the transfer-matrix method and numerical simulations (see Fig. 12).
\begin{figure}
    \centering
    \includegraphics[width=10cm]{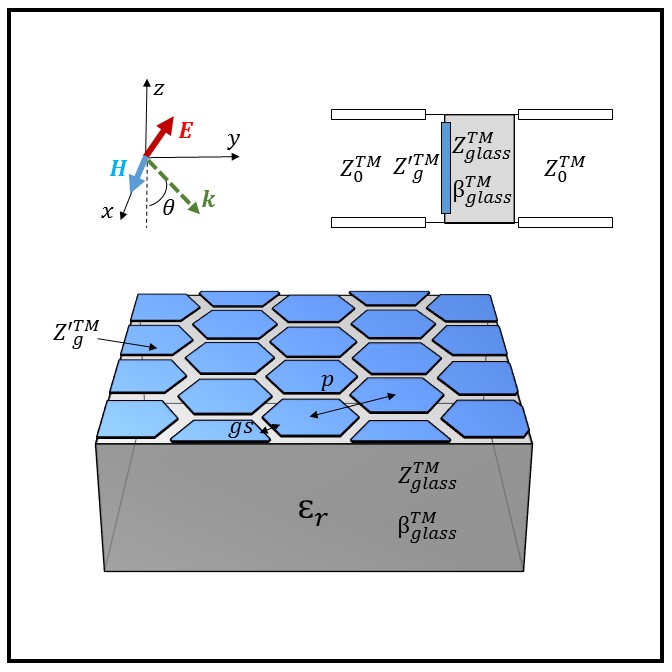}
    \caption{3D schematic and transmission line model of a single honeycomb patch array deposited on one side of the glass substrate}
    \label{fig:f7}
\end{figure}
\begin{figure}
    \centering
    \includegraphics[width=15cm]{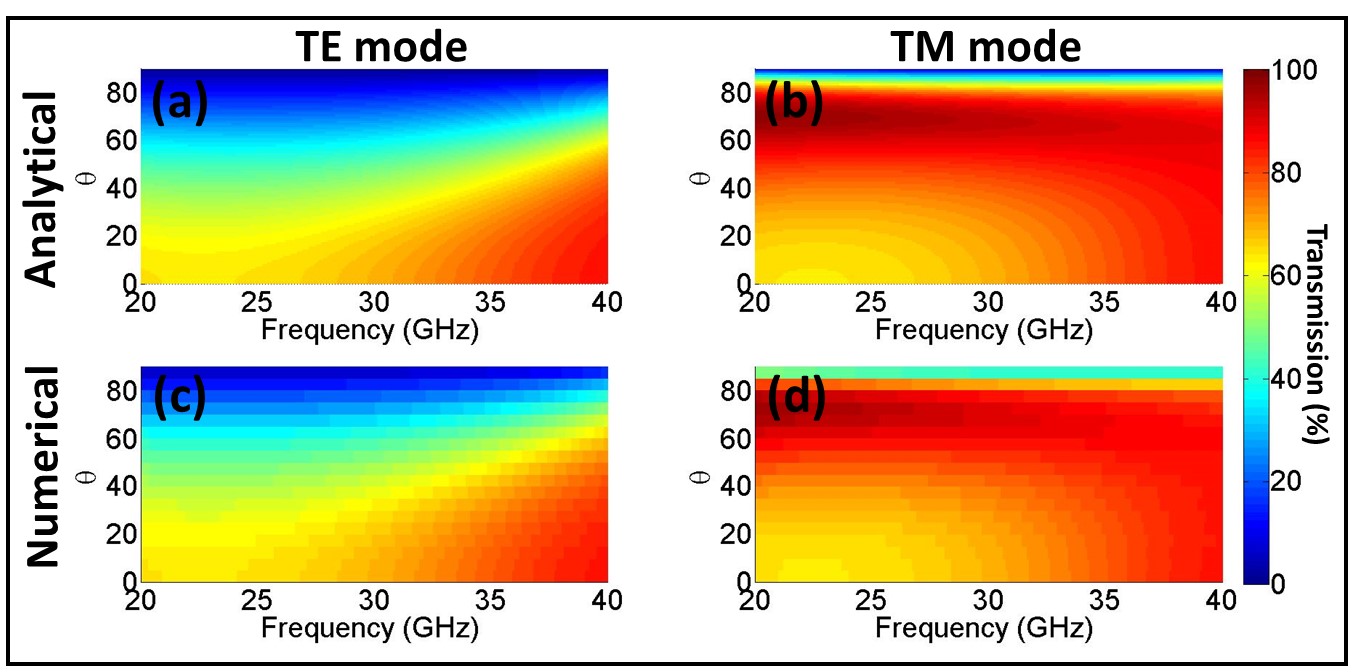}
    \caption{Analytically modelled angular RF transmission under (a) TE- and (b) TM-polarized incident waves, and numerically modelled RF response under (c) TE- and (d) TM-polarized incident waves for a honeycomb patch array ($p=800 \mu m$ and $gs=50 \mu m$) deposited on one side of the glass substrate.}
    \label{fig:f7}
\end{figure}

\section{Angular Performance and Polarization Fidelity: Honeycomb Patch Meta-glass vs Honeycomb Mesh Meta-glass}

In order to achieve wide-angle impedance matching, Safari \textit{et al.} have discussed the need to have both the effective relative permittivity and permeability of the meta-glass equal to unity \cite{Safari2020optically}. However, non-idealities in the fabrication process, the presence of absorptive and scattering losses in the materials\cite{PolyanskiyMN}, and the angular dependency of the effective permittivity associated with the meta-structure lead to non-ideal impedance matching at all incident angles and thus the observed angle-dependent response. 

 \begin{figure}
    \centering
    \includegraphics[width=15cm]{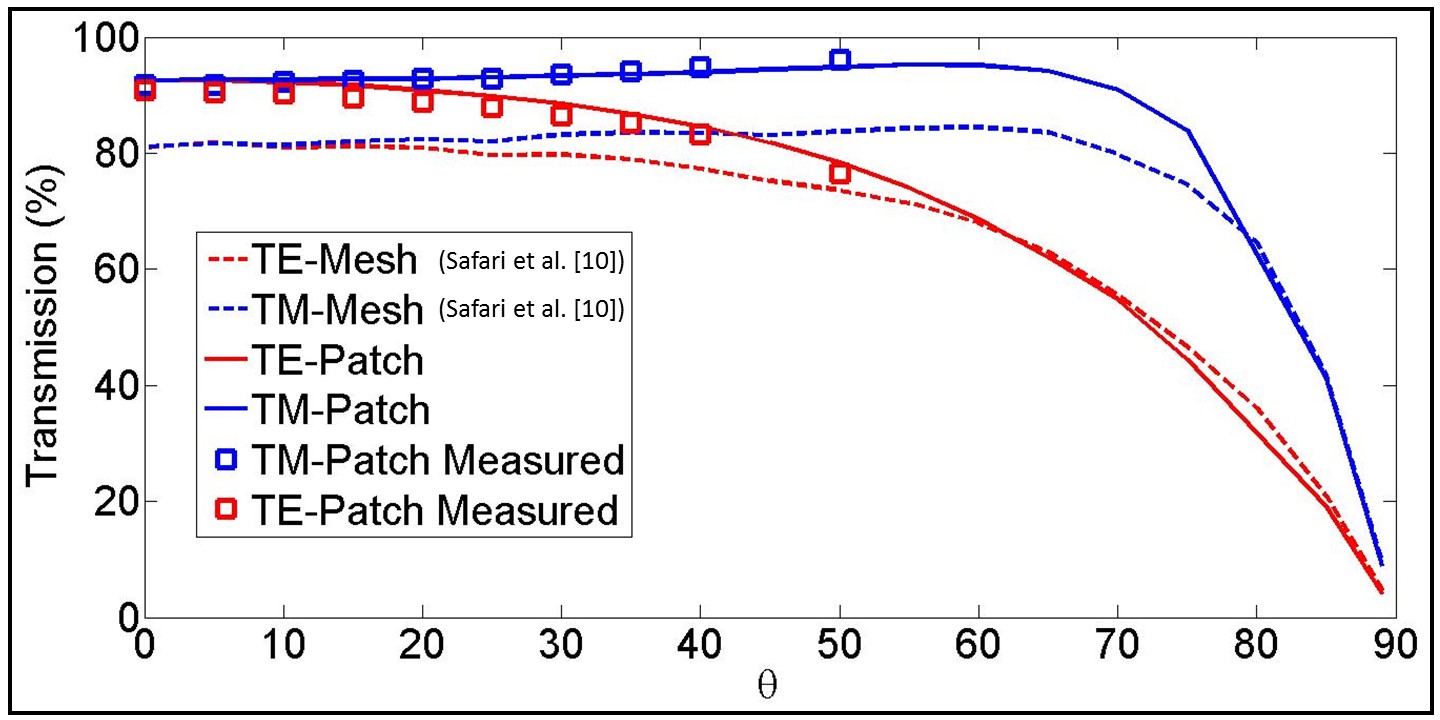}
    \caption{Modelled and measured angular RF transmission of the honeycomb mesh\cite{Safari2020optically} and honeycomb patch meta-glass structures under TE and TM  incident waves.}
    \label{fig:f2}
\end{figure}

Here, we compare the angular performance of the honeycomb patch meta-glass with that of the honeycomb mesh meta-glass, reported by Safari \textit{et al.} in an earlier study \cite{Safari2020optically}, under both TE and TM polarizations. Figure 13 illustrates the angular transmission and polarization fidelity of these two metamaterial structures. We see that the honeycomb patch meta-glass outperforms the honeycomb mesh meta-glass under all incident angles. However, polarization fidelity, that is, consistent performance of the meta-glass for both polarizations under a wide range of incident angles, is slightly better in the case of honeycomb mesh meta-glass. Therefore, the honeycomb patch meta-glass affects the polarization of the incident wave slightly more than the mesh structure. On the other hand, the RF transmission of the patch structure is higher than that of the mesh meta-glass by 10 \% over a wide range of incident angles (0-60$^{o}$).

\
 For the interested reader, a more detailed explanation of the angular performance of composite metasurfaces can be found in \cite{he2019thin}.

\section{Optical Performance: Honeycomb Patch Meta-glass vs Honeycomb Mesh Meta-glass}
\begin{figure}
    \centering
    \includegraphics[width=15cm]{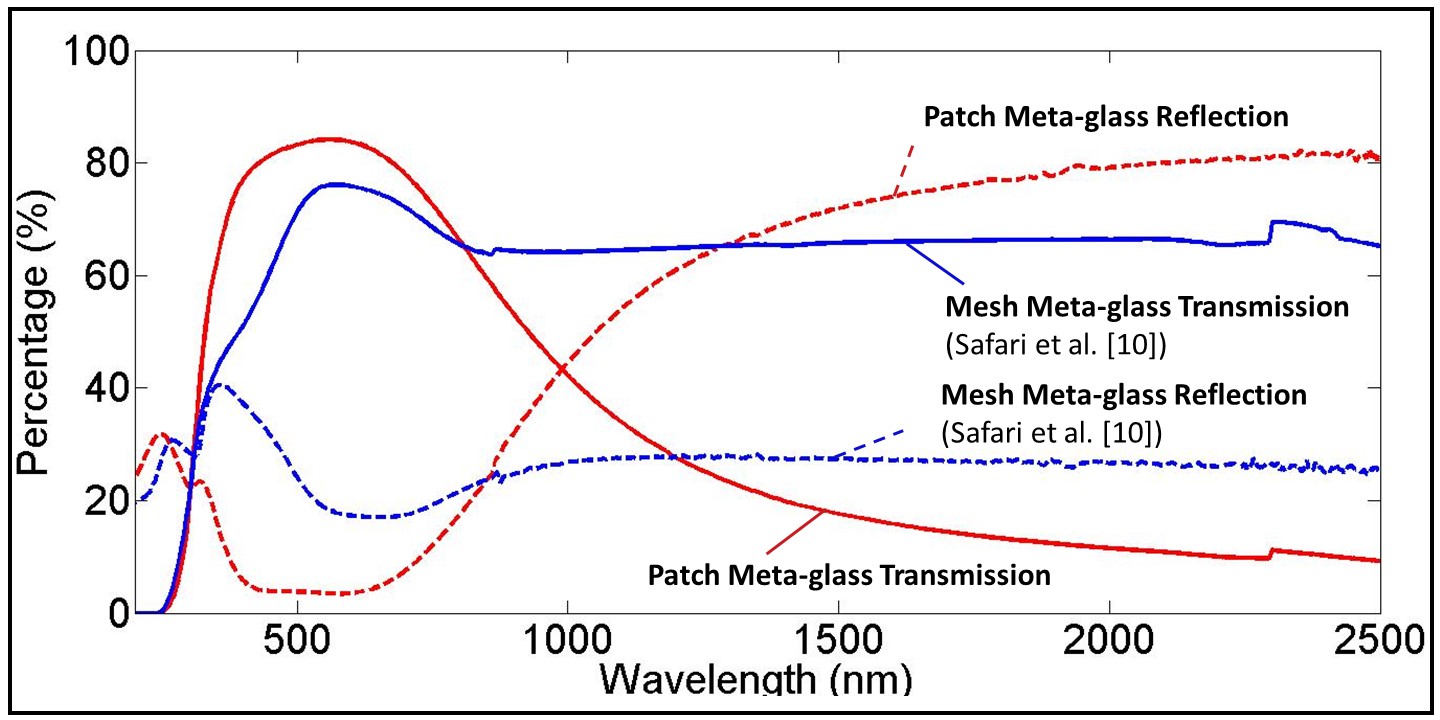}
    \caption{Optical transmission and reflection of the honeycomb patch and honeycomb mesh\cite{Safari2020optically} meta-glasses in the visible and near-IR range}
    \label{fig:f3}
\end{figure}

Here we compare the optical performance of the honeycomb patch meta-glass and honeycomb mesh meta-glass. The main difference between the two structures is that the honeycomb patch meta-glass in the NIR and MIR region outperforms the honeycomb mesh meta-glass. Figure 14 demonstrates the transmission and reflection of both structures in the visible and near-IR. While honeycomb mesh meta-glass exhibits only 28 \% reflection in the near-IR region the honeycomb patch meta-glass outperforms the mesh structure by 50 \% (i.e., exhibiting 80\% reflection). It is interesting to note that the patch metasurface (on each side) consists of only 10 nm Ag while the mesh metasurface consists of 30 nm Ag; admittedly, the patch array effectively functions with a larger surface albeit thinner silver while the mesh grid in contrast requires thicker Ag. We also observe that the patch meta-glass possesses higher peak visible transparency at 550 nm (i.e., 83 \%) when compared to the mesh meta-glass (i.e., 78\%) which is attributed to the thicker Ag. Further, the patch structure appears quite transparent to the average human eye where the lower optical contrast associated with the smaller feature sizes helps with its clarity (see Fig. 15).
\begin{figure}
    \centering
    \includegraphics[width=15cm]{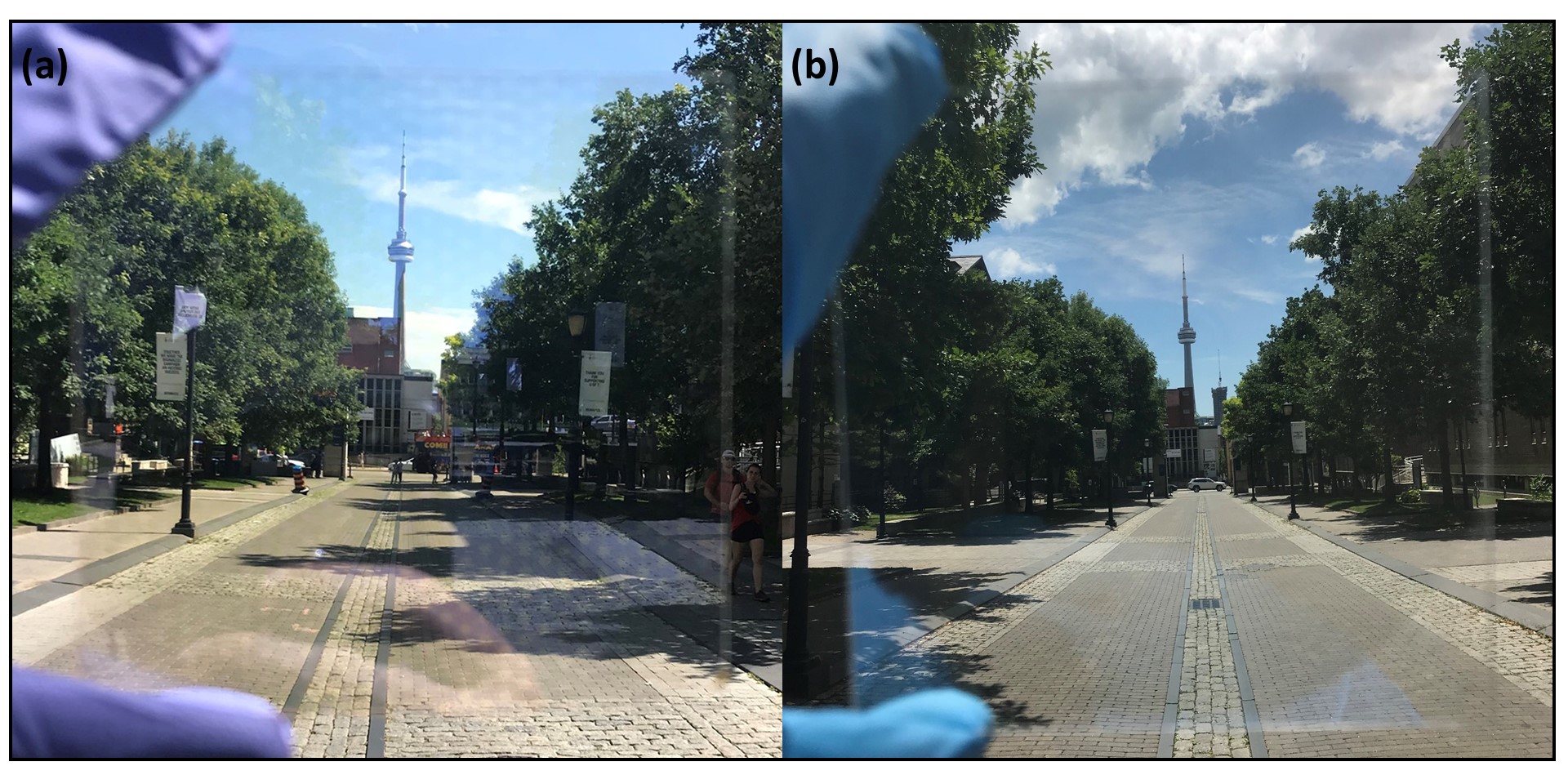}
    \caption{Fabricated honeycomb (a) mesh\cite{Safari2020optically} and (b) patch meta-glasses and their optical contrast.}
    \label{fig:f3}
\end{figure}
\section{Surface Currents}
In order to understand the capability of the honeycomb metasurface in mimicking the performance of the shunt impedances in the transmission-line approach, we demonstrate the induced surficial currents on the honeycomb patch elements (see Fig. 17). Upon closer observation, it is clear that the small sized structure, compared to the operating wavelength, results in a close-to-uniform current distribution. This is another indication that validates the earlier assumption in which the densely packed honeycomb patch array was replaced by a shunt surface impedance element (see Eqn. 5).  
\begin{figure}
    \centering
    \includegraphics[width=15cm]{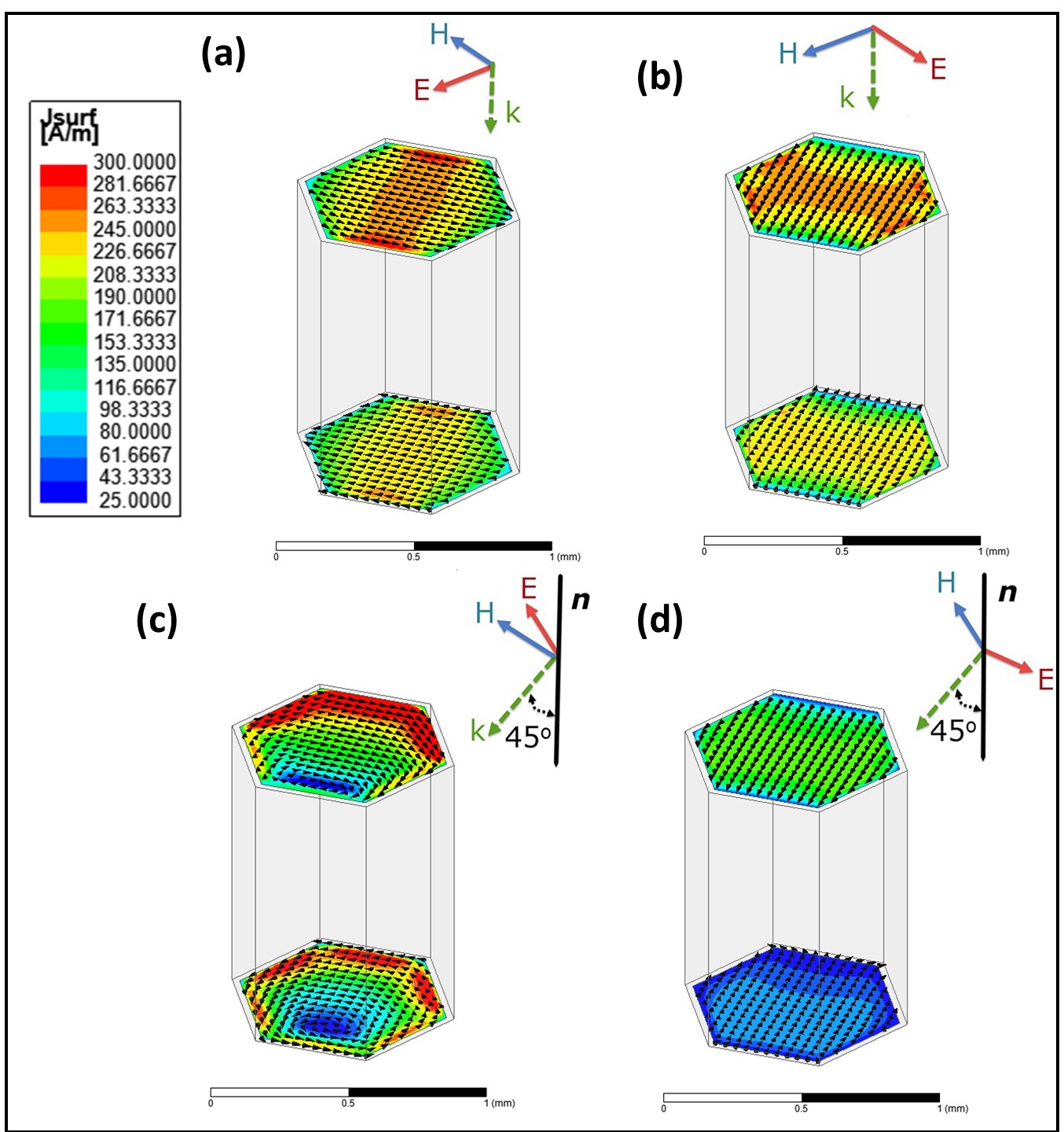}
    \caption{The surface current distribution under (a) TE- and (b) TM-polarized wave at normal angle and under (c) TE- and (d) TM-polarized wave at 45$^o$ angle. }
    \label{fig:f7}
\end{figure}
\end{suppinfo}

\bibliography{achemso-demo}

\end{document}